# Electrical detection of magnetic states in crossed nanowires using the topological Hall effect

Kenji Tanabe[1] and Keisuke Yamada[2]

[1]Department of Physics, Graduate School of Science, Nagoya University, Nagoya, 464-8602, Japan

[2]Department of Chemistry and Biomolecular Science, Faculty of Engineering, Gifu University, Gifu, 501-1193, Japan

**Abstract**

We used micromagnetic simulations to investigate the spatial distributions of the effective magnetic fields induced by spin chirality in crossed nanowires with three characteristic magnetic structures: a radiated-shape, an antivortex, and a uniform-like states. Our results indicate that, unlike the anomalous Hall effect, the topological Hall effect (which is related to the spin chirality) depends on both the polarity and the vorticity. Therefore, measuring the topological Hall effect can detect both the polarity and the vorticity simultaneously in crossed nanowires. This approach may be suitable for use as an elemental technique in the quest for a next-generation multi-value memory.

**Introduction**

Recently, non-uniform and nanometer-scale magnetic structures such as magnetic domain walls,[1–5] magnetic vortices,[6–12] antivortices,[13–16], and skyrmions[17–21] have been attracting increasing attention for use as high-density memory cells for future non-volatile data-storage devices because of their smallness and their magnetic stability at room temperature. The magnetic vortex and antivortex are one of the most fundamental magnetic states, with two degrees of freedom, namely a polarity and a vorticity. The polarity ($P$) determines the direction of the perpendicular component of magnetization at the center of a vortex state, the region of which is called a core; an up (down) core is denoted as $P = +1(-1)$. The vorticity ($V$) characterizes the vortex winding structure, which consists of the in-plane components and is defined by

$$V = \oint \nabla \times \boldsymbol{n} \cdot d\boldsymbol{r}, \qquad (1)$$

where $\boldsymbol{n}$ is the unit vector of the magnetization, and the integral range is an arbitrary closed loop around the core. A magnetic structure with $V = +1(-1)$ is called a magnetic vortex (antivortex). Electrical control and detection are crucial if these degrees of freedom are to be applied to a magnetic memory cell. In previous studies, magnetic-domain-wall motion,[2–5] core switching in the magnetic vortex,[8–10] and skyrmion dynamics[20, 21] induced by spin torque have been demonstrated as electrical control methods. In contrast, the magnetic states have often been observed by non-electrical methods such as magnetic-force microscopy,[2, 4, 6, 8, 9, 13, 15, 16] spin-polarized scanning tunneling microscopy,[7] Lorentz-type scanning electron microscopy,[19] and magneto-optic Kerr-effect microscopy.[3] Although there have been some studies on electrical detection by giant magnetoresistance,[1] tunneling magnetoresistance,[10] and anisotropic magnetoresistance,[5, 11, 12, 16] only dynamical methods such as magnetic resonance[12] are capable of detecting both the polarity and vorticity simultaneously. Hence, other methods for the electrical detection of these degrees of freedom are required.

As a method for electrical detection, we consider the Hall effect induced by the (scalar) spin chirality, which is called the topological Hall effect (THE). The spin chirality is defined by

$$\sum \bm{s}_i \cdot (\bm{s}_j \times \bm{s}_k), \quad (2)$$

where $\bm{s}_i$, $\bm{s}_j$, and $\bm{s}_k$ denote three non-coplanar spins.[22] Spin chirality creates effective magnetic and electric fields only for spin on a conduction electron. This differs from the external magnetic and electric fields, and is attracting significant interest as an emergent electromagnetic phenomenon.[22–24] This effective magnetic field has been detected experimentally as THE in pyrochlore crystals[25-27] and skyrmion lattices.[28, 29] On the other hand, the effective electric field arises in non-uniform magnetic dynamics such as magnetic-domain-wall motion[30, 31] and magnetic-vortex dynamics[32] as a spin-motive force.[33-35] Since spin chirality exists in non-uniform magnetic states, THE measurements may be suitable for electrical detection of magnetic vortices and antivortices.

The magnetic states are determined from the condition wherein the total magnetic energy (including the magneto-static energy and magneto-exchange interaction) is minimized. In nanometer-scale strips, uniform magnetic structures are the norm because the magneto-exchange interaction is much larger than the other interactions in most ferromagnetic materials. Non-uniform magnetic states appear in unique nanostructures such as a disk or a cross, in which the magneto-static energy is enhanced by geometrical effects. A pair of nanowires shaped as a cross is a rare system wherein magnetic states with three different vorticities can be stable, as shown in Fig. 1. The vorticity in the crossed nanowires is assigned by the directions of the magnetizations in the four nanowire arms. The magnetic states with $V = +1, 0, -1$ are referred to as radiated-shape, uniform-like, and antivortex states, respectively (Fig. 1(b)). In the present letter, we investigate the spatial distributions of the effective magnetic fields induced by spin chirality in crossed nanowires with various magnetic states. The results show that, unlike the anomalous Hall effect (AHE), THE depends on both polarity and vorticity.

**Methods**

A two-dimensional micromagnetic model was used in the simulation.[36] The thickness ($D$), length ($L$), and width ($W$) of the crossed nanowires were 10 nm, 256 nm, and 10–48 nm, respectively, as shown in Fig. 1(a). The structure was divided into rectangular prisms with dimensions of $1 \times 1 \times$ 10 nm³. The typical material parameters of a permalloy (Ni$_{81}$Fe$_{19}$) were used, as follows: exchange stiffness constant $A = 1 \times 10^{-6}$ erg/cm, saturation magnetization $M_s = 800$ emu/cm³, magnetocrystalline anisotropy $K_u = 0$, and Gilbert damping constant $\alpha = 1$. Here we used a relatively high damping in order to explore the static magnetic states. Using the equilibrium magnetic states, we calculated the effective magnetic field from

$$B_z^{\text{eff}} = -\frac{h}{8\pi e} \mathbf{n} \cdot \left(\frac{\partial \mathbf{n}}{\partial x} \times \frac{\partial \mathbf{n}}{\partial y}\right), \quad (3)$$

where $h$ is Planck's constant, $e$ is the charge on an electron, and $\mathbf{n}$ is the unit vector of the magnetization. In this letter, we represent an effective magnetic field as $B_z^{\text{eff}}$ (G) and use the centimeter–gram–second system of units.

**Results and discussion**

Figures 2(a–c) illustrate the spatial distributions of the radiated-shape, uniform-like, and antivortex magnetic states in the crossed nanowires with $W = 48$ nm at equilibrium. A magnetic vortex structure appears in the crossover area in the radiated-shape state (Fig. 2(a)). This is because formation of the vortex state suppresses an increment in the magneto-static energy due to concentration of magnetic poles. In contrast, a vortex state does not appear in narrower nanowires of width $W \leq 14$ nm (data not shown), which suggests that the magneto-exchange interaction becomes more dominant than the magneto-static interaction. Both the radiated-shape and vortex states have the same vorticity, $V = +1$, and they are classified into the same group. We can also find a core at the center of the vortex, as shown in Fig. 2(a). Figure 2(b) shows the antivortex state; a core that is similar to the one in the radiated-shape state appears at the center of the antivortex. The uniform-like state does not have perpendicular components of magnetization, as shown in Fig. 2(c).

Figures 2(d–f) show the spatial distributions of the effective magnetic fields, $B_z^{\text{eff}}$, calculated from the magnetic states in Figs. 2(a–c). Non-zero magnetic fields are induced in both the radiated-shape and antivortex states in Figs. 2(d,e). Since the polarities have the same sign, the effective magnetic fields have opposite signs. In contrast, the uniform-like state produces few magnetic fields, a feature that is due to the appearance of spin chirality in only the non-uniform state.

Let us consider the external-magnetic-field dependence of the magnetic states and effective magnetic fields in Fig. 3. The core sizes depend strongly on the external magnetic field. The core size increases with increasing external field parallel to the core direction in both the radiated-shape and antivortex states. In contrast, whereas the signs of the effective magnetic field, $B_z^{\text{eff}}$, in the radiated-shape and antivortex states with $P = +1$ are opposite, the maximum value of the magnitude of the effective magnetic field, $B_z^{\text{eff}}$, increases with decreasing external field and exceeds 180 kG at $H_z^{\text{ext}} = -4$ kOe. This appreciable enhancement is because the effective magnetic flux concentrates at the center of the crossover area because of the decrease in core size and the fact that negative perpendicular components appear around the core in the range of $H_z^{\text{ext}} < 0$.

Next, we argue for the Hall effects in the crossover area. The Hall effects in ferromagnetic materials consist of the ordinary Hall effect (OHE), the planar Hall effect, AHE, and THE. Here, we ignore the planar Hall effect because we only discuss the dependence on perpendicular external fields. Accordingly, the Hall resistivity, $\rho_{xy}$, is expressed by

$$\rho_{xy} = \frac{1}{S}(R_o \Phi^{\text{ext}} + R_a \Phi^{\text{Mz}} + R_o \Phi^{\text{Bz}}), \quad (4)$$

where $S$ is the crossover area, $R_o$ is the Hall coefficient due to OHE and THE, $\Phi^{\text{ext}}$ is the external magnetic flux in the crossover area, $R_a$ is the Hall coefficient due to AHE, $\Phi^{\text{Mz}}$ is the integral value of the perpendicular component of the magnetization in the crossover area, and $\Phi^{\text{Bz}}$ is the effective magnetic flux in the crossover area. The second term on the right-hand side of Eq. (4) is ascribed to AHE and is proportional to the perpendicular component of the magnetization. The third term is ascribed to THE and depends on the effective magnetic flux. According to previous study [37], $R_o$ and $R_a$ in thin epitaxial $Ni_{80}Fe_{20}$ films with 90-nm thickness are approximately $10^{-12}$ Ωcm/G at room temperature, which corresponds to a carrier density of $10^{23}$ cm$^{-3}$. Thus, when $\Phi^{\text{Mz}} \ll \Phi^{\text{Bz}}$ and $\Phi^{\text{ext}} \ll \Phi^{\text{Bz}}$, the THE contribution becomes dominant in Hall measurements.

Figure 4(a) shows the external-field dependence of $\Phi^{\text{Mz}}$ and $\Phi^{\text{ext}}$ in the crossed nanowires with $W = 48$ nm. The sign of $\Phi^{\text{Mz}}$ depends only on the polarity near the zero external field. The perpendicular component of the magnetization increases monotonically with increasing external field. The kink in the data at $H_z^{\text{ext}} = 7$ kOe for the radiated-shape state with $P = +1$ corresponds to core annihilation. The data for the antivortex state with $P = -1$ increase rapidly at $H_z^{\text{ext}} = 6$ kOe and merge with those for the antivortex state with $P = +1$, which corresponds to core switching. Conversely, core switching does not occur in the radiated-shape state up to core annihilation, which suggests that the radiated-shape state is more stable than the antivortex state. Since the data for the radiated-shape and antivortex states with the same polarity are quite similar, we cannot distinguish the vorticity from the anomalous Hall measurements.

Finally, we discuss the external-field dependence of the effective magnetic flux, $\Phi^{\text{Bz}}$, as shown in Fig. 4(b). Because the topological Hall signal detected in the nanowires is proportional to the effective magnetic flux (as described by Eq. (4)), Fig. 4(b) corresponds to the external-field dependence of THE. The magnitudes of the effective magnetic fluxes are independent of both the polarity and vorticity near the zero external field; they are approximately the quantized magnetic flux, $h/4e \sim 2 \times 10^{-7}$ (Mx). The reason for this can be understood by integrating Eq. (3):

$$\Phi^{\text{Bz}} = -\frac{h}{4e} \iint \left\{ \frac{1}{2\pi} \boldsymbol{n} \cdot \left( \frac{\partial \boldsymbol{n}}{\partial x} \times \frac{\partial \boldsymbol{n}}{\partial y} \right) \right\} dxdy. \quad (5)$$

The magnitude of the integral on the right-hand side of Eq. (5) is unity in the radiated-shape and antivortex states; its sign is determined by the product of the polarity and vorticity, $PV$. Hence, THE does not depend only on the polarity, but also on the in-plane components of the magnetizations. According to previous studies on skyrmions, twice the integral on the right-hand side of Eq. (5) is known as the skyrmion number. Furthermore, the external-field dependence of the effective magnetic flux is quite different from that of the magnetization as shown in Fig. 4(a). Even though the effective magnetic fluxes in the radiated-shape state with $P = -1$ and the antivortex state with $P = +1$ are the same in the absence of a magnetic field, they increase and decrease with increasing external magnetic field in the radiated-shape state with $P = -1$ and in the antivortex state with $P = +1$, respectively. This is because the magnitude of the integral, which is related to the skyrmion number, now differs from unity because of the tilted magnetization in the external perpendicular field. As stated already, the ordinary and anomalous Hall coefficients in thin epitaxial $Ni_{80}Fe_{20}$ films are of the same order of magnitude. Since $\Phi^{Mz} \ll \Phi^{Bz}$ and $\Phi^{ext} \ll \Phi^{Bz}$ under the low magnetic field ($|H^{ext}| < 1$ kOe) in Fig. 4, the THE contribution becomes dominant in Hall measurements. The sign and field dependence of the Hall voltage under the low magnetic field provides information about both the polarity and the vorticity. Moreover, the THE contribution becomes more dominant in narrower wires because OHE and AHE are proportional to the crossover area and THE is determined by Eq. (5).

## Conclusion

We investigated the spatial distributions of the magnetic states and the effective magnetic fields induced by spin chirality in the crossed nanowires using micromagnetic simulations. The crossed nanowires have three characteristic magnetic structures with different vorticities, and the effective magnetic fields depend strongly on vorticity. Whereas the effective magnetic field is antiparallel to the external magnetic field in the radiated-shape magnetic state with $V = +1$, it is parallel to the external magnetic field in the antivortex magnetic state with $V = -1$. In contrast, no effective magnetic field is induced in the uniform-like magnetic state with $V = 0$. The external-field dependence depends on both polarity and vorticity. Therefore, these results reveal that Hall measurements can detect both polarity and vorticity simultaneously in crossed nanowires. This approach may be suitable for use as an elemental technique in the quest for a next-generation multi-value memory.

## Acknowledgments

We would like to thank Prof. Y. Nakatani for collaboration at an early stage of this work. This work was partly supported by a Grant-in-Aid for Challenging Exploratory Research (No.15K13272) from the Japan Society for the Promotion of Science (JSPS), a Grant-in-Aid for Young Scientists (B)

Figure 1

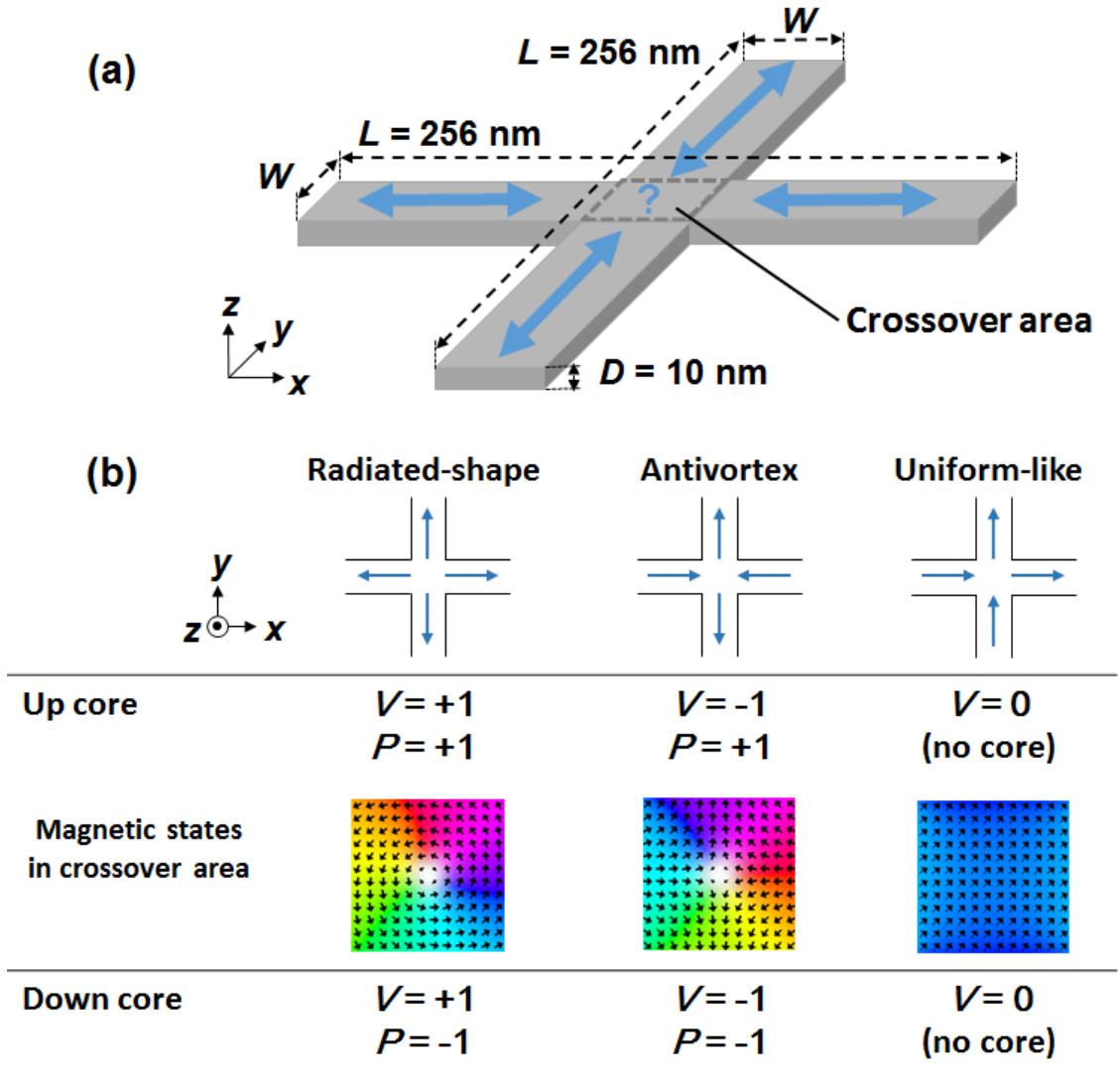

FIG 1(a) Schematic diagram of crossed nanowires. The nanowires have four arms, and the blue allows indicate the directions of the magnetizations. The directions are parallel (antiparallel) to the nanowires because of a suppressed magneto-static energy. The characteristics of the magnetic structure in the crossover area depend on the direction of the arm magnetizations. (b) Relationships between magnetic state, vorticity, and polarity. The magnetic states are classified into radiated-shape, antivortex, and uniform-like states by the blue allows in the first row in the table. The radiated-shape, antivortex, and uniform-like states have vorticities of $V = +1$, $V = -1$, and $V = 0$, respectively. The magnetic states in the crossover area are shown in the color plots in the second row. Whereas the uniform-like state does not have a core, the radiated-shape and antivortex states have up and down cores, which are denoted as $P = +1$ and $P = -1$, respectively.

**Figure 2**

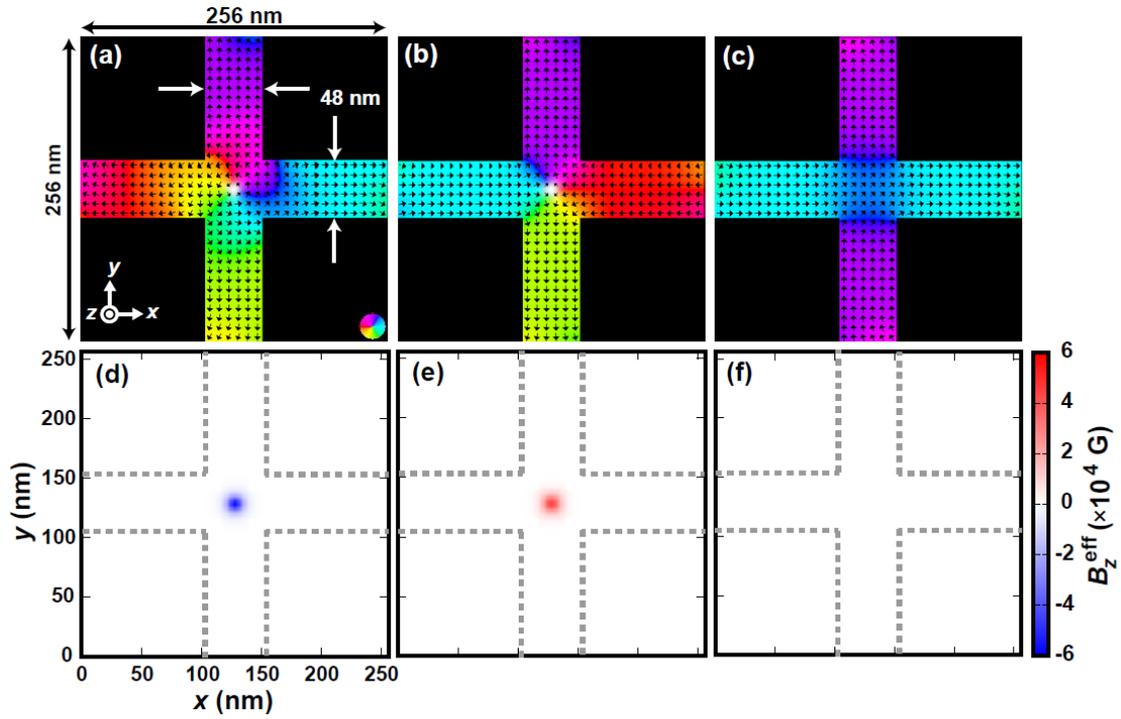

FIG 2 Spatial distributions of the magnetizations (a–c) and effective magnetic fields (d–f) in the radiated-shaped, uniform-like, and antivortex states in crossed nanowires with $W = 48$ nm. The colors in (a–c) indicate the in-plane components of the magnetizations. The gray dotted lines in (d–f) denote the region of the nanowires.

**Figure 3**

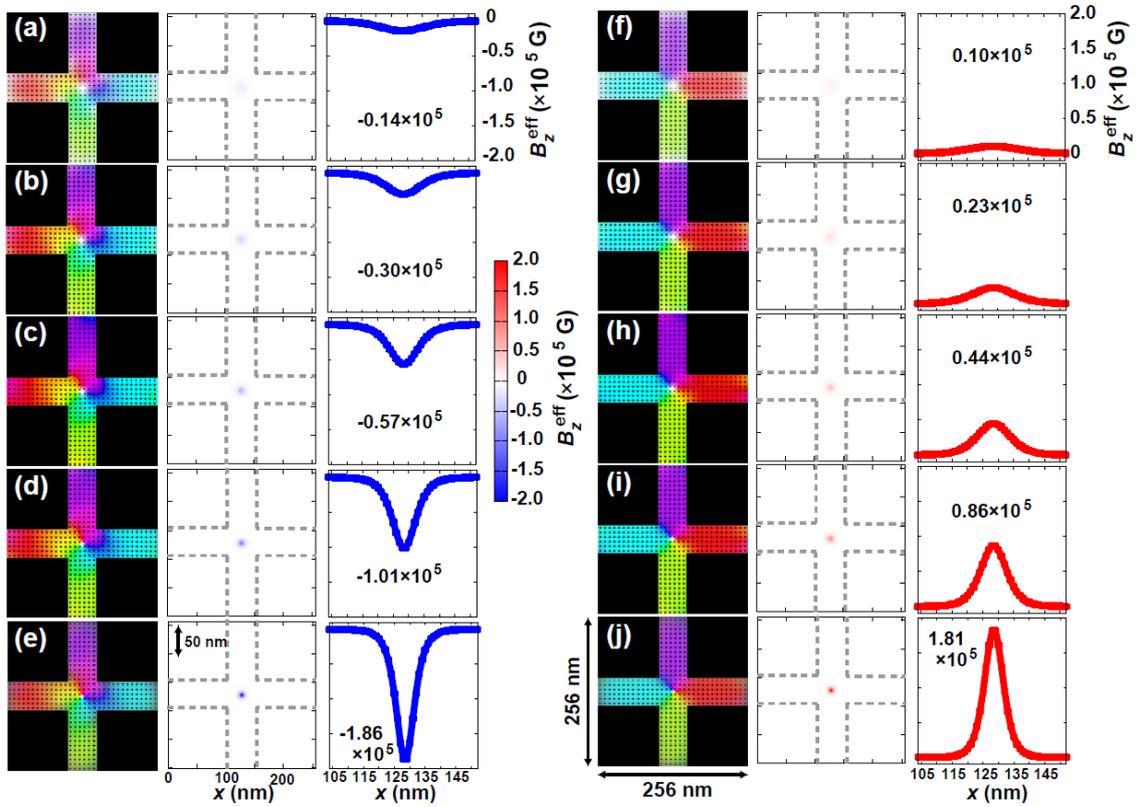

FIG 3 Magnetic states, effective-magnetic-field distributions, and their profiles in the crossover area in the radiated-shape (a–e) and antivortex (f–j) states with $P = +1$ in an external perpendicular magnetic field $H_z^{ext} = +4\,\text{kOe}$ (a,f), $+2\,\text{kOe}$ (b,g), $0\,\text{kOe}$ (c,h), $-2\,\text{kOe}$ (d,i), and $-4\,\text{kOe}$ (e,j). Values in the profiles denote the effective magnetic field (G) at the center.

**Figure 4**

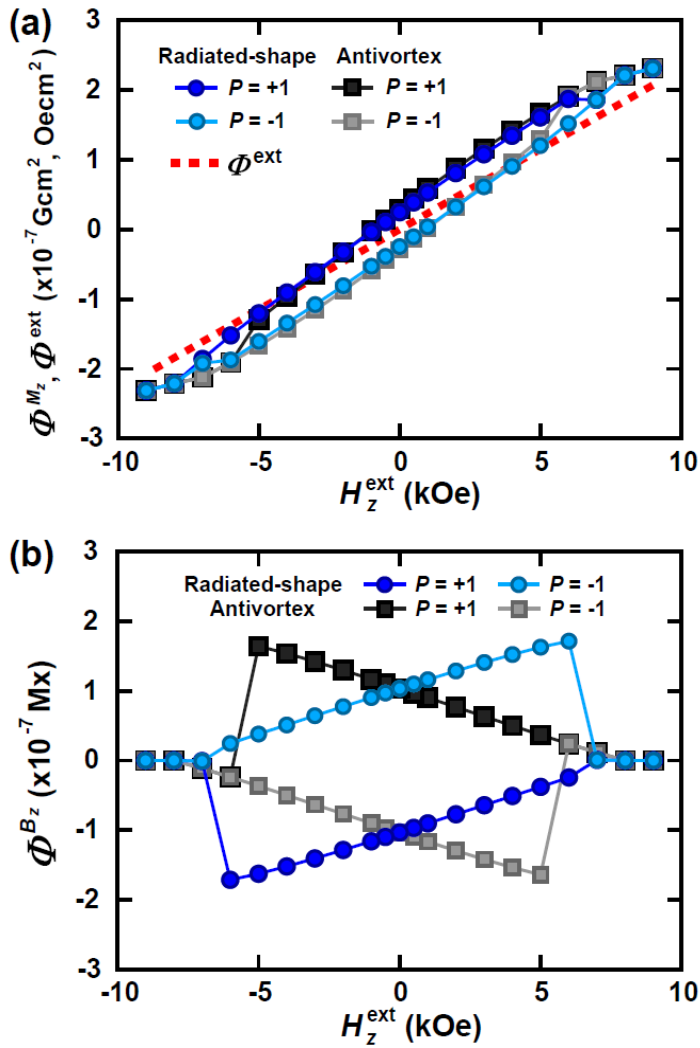

FIG 4(a) External-magnetic-field dependence of $\Phi^{\text{ext}}$ and $\Phi^{M_z}$, which is the integral value of the perpendicular component of the magnetization in the crossover area in crossed nanowires with $W = 48$ nm. (b) External-magnetic-field dependence of the effective magnetic flux, $\Phi^{B_z}$, which is the integral value of the perpendicular component of the effective magnetic field in the crossover area. Dimension of $\text{Mx}$ is equal to that of $\text{Gcm}^2$ and $\text{Oecm}^2$.